\documentclass[prd,amsfonts,amssymb,noshowpacs,nofootinbib,preprintnumbers]{revtex4}
\usepackage{amsmath,graphicx}

\begin{document}

\preprint{YITP-12-33}
\title{Strong scale dependent bispectrum in the Starobinsky
model of inflation}
\author{Frederico Arroja$^{1}$\footnote{arrojaf{}@{}ewha.ac.kr}
and Misao Sasaki$^{2}$\footnote{misao{}@{}yukawa.kyoto-u.ac.jp}
}
\affiliation{
{}$^{1}$Institute for the Early Universe,
Ewha Womans University, Seoul 120-750, Republic of Korea
\\
{}$^{2}$Yukawa Institute for Theoretical Physics,
Kyoto University, Kyoto 606-8502, Japan
}

\begin{abstract}
We compute analytically the dominant contribution to the tree-level
 bispectrum in the Starobinsky model of inflation. In this model,
the potential is vacuum energy dominated but contains a subdominant
linear term which changes the slope abruptly at a point.
We show that on large scales compared with the transition scale $k_0$
and in the equilateral limit the analogue of the non-linearity parameter
 scales as $(k/k_0)^2$, that is its amplitude decays for larger and
larger scales until it becomes subdominant with respect to the usual
slow-roll suppressed corrections.
On small scales we show that the non-linearity parameter oscillates with
angular frequency given by $3/k_0$ and its amplitude grows linearly towards
smaller scales and can be large depending on the model parameters.
 We also compare our results with previous results in the literature.
\end{abstract}


\date{\today}
\maketitle

\section{Introduction\label{sec:introduction}}

The idea of inflation in the very early
universe~\cite{Starobinsky:1979ty,Kazanas:1980tx,Sato:1980yn,Guth:1980zm,Linde:1981mu,Albrecht:1982wi} is a very compelling solution to the problems of
the Big Bang theory. However one should keep in mind that there may
be other viable alternative explanations. Other successful
predictions~\cite{Komatsu:2010fb} of the simplest inflationary models are
the nearly scale invariant, mostly of adiabatic origin and nearly Gaussian
 primordial curvature perturbation. This primordial perturbation is generated
due to quantum vacuum fluctuations of the scalar field driving inflation
and will at a much later stage in the evolution of the universe give
origin to the cosmic microwave background (CMB) radiation anisotropies and
 the large-scale structure of galaxies in the universe.

While the simplest models are appealing due to their simplicity and fewer
ingredients they are not always realistic and often more realistic models
introduce several fields, non-trivial potentials and kinetic terms, etc..
All these models have to pass the observational constraints imposed by
 current data on the cosmological background as well as on the power
spectrum of the perturbations. In many cases the lack of other observables
renders many models observationally indistinguishable. However some models
have other observables like non-Gaussianity (i.e. non-linearity
of the perturbations) and these can be and are being used to further
 constrain and distinguish models.

Non-Gaussianity of the primordial curvature perturbation has not been
 observed, however the CMB data still allows a small non-Gaussianity
($\sim 0.1\%$) and there have been some claims of detections or hints of
its presence~\cite{Yadav:2007yy,Xia:2010yu,Hoyle:2010ce,Enqvist:2010bg}.
Despite this small value, efforts are underway (e.g. ESA's Planck
satellite~\cite{PLANCK}) to measure the CMB anisotropies to a precision
with which this small amount would be detected to many standard deviations.

It also turns out that the simplest models of inflation, i.e. the ones
driven by a single scalar field with standard kinetic term, with a potential
 satisfying the slow-roll conditions and if one assumes the standard initial
conditions for the quantum vacuum fluctuations, predict a small and
unobservable level of primordial non-linearity~\cite{Maldacena:2002vr,Acquaviva:2002ud,Seery:2006vu,Seery:2008ax,Seery:2006js,Byrnes:2006vq}.
Therefore only models that can produce large non-Gaussianity can be
constrained if no detection is made. One of such models put forward by
Starobinsky in 1992~\cite{Starobinsky:1992ts} will be the focus of this work.
Many models which break at least one of the above conditions have been
proposed and studied recently. An incomplete list is:
for models that use several fields \cite{Dvali:2003em,Enqvist:2004ey,Lyth:2005qk,Lyth:2005fi,Alabidi:2006hg,Sasaki:2006kq,Valiviita:2006mz,Sasaki:2008uc,Naruko:2008sq,Suyama:2008nt,Byrnes:2008wi,Byrnes:2008zz,Byrnes:2008zy,Cogollo:2008bi,Rodriguez:2008hy,Gao:2008dt,Langlois:2008vk,Langlois:2008wt,Langlois:2008qf,Arroja:2008yy,Chen:2009zp,Huang:2009xa,Huang:2009vk,Byrnes:2009qy,Langlois:2009ej,Mizuno:2009cv,Mizuno:2009mv,Gao:2009at,RenauxPetel:2009sj,Cai:2009hw,Kim:2010ud,Gao:2010xk},
for models with non-canonical kinetic terms \cite{Creminelli:2003iq,Alishahiha:2004eh,Gruzinov:2004jx,Chen:2006nt,Huang:2006eha,Arroja:2008ga,Chen:2009bc,Arroja:2009pd,Huang:2010ab,Izumi:2010wm,Mizuno:2010ag,Burrage:2010cu}, models with temporary violations of the slow-roll conditions or small departures of the initial vacuum state from the standard Bunch-Davies vacuum \cite{Chen:2006xjb,Chen:2008wn,Hotchkiss:2009pj,Hannestad:2009yx,Flauger:2010ja,Chen:2010bka,Takamizu:2010xy,Agullo:2010ws,Ganc:2011dy,Park:2012rh}. For reviews on the theory of non-Gaussian perturbations see \cite{Koyama:2010xj,Chen:2010xk,Tanaka:2010km,Byrnes:2010em,Wands:2010af}.

The initial motivation of the Starobinsky model~\cite{Starobinsky:1992ts}
that breaks temporarily the slow-roll approximation was to explain
 an observation that the correlation function of galaxies seemed to require
 more power on large scales than at smaller length scales within the paradigm
of a Einstein-de-Sitter universe (a spatially flat universe with cold dark
matter particles). It was also used to explain the possible nonconservation
(enhancement) of the curvature perturbation on superhorizon scales due
to not-fast-enough decay of the decaying mode~\cite{Leach:2001zf}.
More recently similar
models~\cite{Starobinsky:1992ts,Adams:2001vc,Gong:2005jr,Joy:2007na,Biswas:2010si,Nakashima:2010sa,Chen:2011zf} were
revived and proposed because it was realized that they can provide better
fits to the CMB anisotropies power spectrum than the primordial
power-law model~\cite{Covi:2006ci,Hamann:2007pa,Joy:2008qd,Mortonson:2009qv,Hazra:2010ve,Adshead:2011jq}.
Another reason for this interest is that it was realized that such
models have very distinctive bispectrum signatures like for instance
 non-linearity parameters that are strongly scale-dependent. A particularly
 appealing feature of the Starobinsky model, a vacuum energy dominated
potential added with a linear term that changes slope abruptly at a point,
is that despite breaking a slow-roll condition it admits an analytical
treatment of the background and linear perturbations equations as well as
the non-Gaussianity as was shown in~\cite{Takamizu:2010xy,Martin:2011sn}
 and in the present paper.

The main goal of this work is to compute analytically the leading-order
bispectrum for the Starobinsky model. The bispectrum on large scales was
first computed in \cite{Takamizu:2010xy} using a next-to-leading order
gradient expansion formulation and more recently on all scales in the
 equilateral limit in \cite{Martin:2011sn}, we compare the results
 of our calculation with these works.

This paper is divided in four sections and an appendix.
 In Section \ref{sec:model}, we describe the model and the analytical
 solutions for the background and linear perturbations are briefly
reviewed. In Section \ref{sec:bispectrum}, we compute the
leading-order contribution to the bispectrum in the equilateral limit.
In the second subsection we shall discuss and compare our results with
 previous attempts in the literature. In the last subsection we
elaborate on the size and shape of the analogue of the non-linearity
parameter in this model. Section \ref{sec:conclusion} is devoted to
 conclusions. In Appendix \ref{FullBispectrum} we present the
lengthy result for the dominant contribution to the bispectrum that is
valid for any triangular configuration (not only in the equilateral limit)
 of the three momenta on which the bispectrum depends.

We work in units where the reduced Planck constant $\hbar$ and the speed
of light $c$ are $\hbar=c=1$ and the reduced Planck mass is
$M_{Pl}=(8\pi G)^{-1/2}$, where $G$ is Newton's gravitational constant.

\section{The model\label{sec:model}}

In this section we introduce the model under consideration and review
the analytical solutions for the background and the perturbation equations.
 All the results of this section can be found in
Refs.~\cite{Starobinsky:1992ts,Martin:2011sn}, here we only briefly
review the main results that will be needed in the computation of
the bispectrum of the next section.
We use a somewhat different notation at points.

The Starobinsky model consists of a canonical scalar field minimally
coupled to Einstein gravity with a potential given by
\begin{eqnarray}
V(\phi)=\left\{
                \begin{array}{lr}
                   V_0+A_+\left(\phi-\phi_0\right) & : \phi > \phi_0,\\
                   V_{0}+A_-\left(\phi-\phi_0\right) & : \phi < \phi_0,
                \end{array}
         \right.\label{potential}
\end{eqnarray}
where $\phi$ is the scalar field value, $V_0$, $A_+$ and $A_-$ are model
parameters that are assumed to be positive. The sharp change in the
slope occurs at $\phi=\phi_0$ and $\phi_0$ is another free parameter
of the model. We also assume that $|V-V_0|/V_0\ll1$, which means that
 the potential is vacuum energy dominated.

In the present case we are interested in a
Friedmann-Lema\^{\i}tre-Robertson-Walker flat, homogeneous and
isotropic background spacetime given by
\begin{equation}
ds^2=-dt^2+a^2(t)\delta_{ij}dx^idx^j,
\end{equation}
where $t$ denotes cosmic time and $a(t)$ is the cosmic scale factor.
The relevant equations of motion are the Friedmann and Klein-Gordon
equations given respectively by
\begin{eqnarray}
H^2=\frac{1}{3M^2_{Pl}}\left(\frac{1}{2}\dot\phi^2+V(\phi)\right),\qquad
\ddot\phi(t)+3H\dot\phi(t)+\frac{dV}{d\phi}=0,
\end{eqnarray}
where the Hubble rate is defined as usual $H=\dot a/a$ and dot denotes
 the derivative with respect to $t$.
The slow-roll parameters are defined by
\begin{equation}
\epsilon=-\frac{\dot H}{H^2}=\frac{\dot\phi^2}{2H^2M^2_{Pl}}, \quad
\eta=\frac{\dot \epsilon}{\epsilon H}
=2\left(\frac{\ddot\phi}{\dot\phi H}+\epsilon\right),
\end{equation}
and will be assumed to be small before the transition.
After the transition $\epsilon$ will continue to be small but
$\eta$ will become large for some time before the slow-roll regime
is again recovered.

With the above mentioned assumption of a vacuum dominated potential the
 dynamics for the scale factor becomes particularly simple and is described by
\begin{equation}
H^2\approx\frac{V_0}{3M^2_{Pl}},
\end{equation}
which has a solution $a(t)=\exp(H_0t)$, where $H_0=\sqrt{V_0/(3M^2_{Pl})}$
 and we chose $a(t=0)=1$. In conformal time, denoted by $\tau$, the previous
 solution is $a(\tau)=-1/(H_0\tau)$ and one has $-H_0t=\ln(-H_0\tau)$.
Before the transition the Klein-Gordon equation can also be easily solved
 to find \cite{Starobinsky:1992ts,Martin:2011sn}
\begin{equation}
\phi_+\simeq\phi_i+\frac{A_+M^2_{Pl}}{V_0}\ln(-H_0\tau),
\qquad \frac{d\phi_+}{d\tau}\simeq-aH\frac{A_+M_{Pl}^2}{V_0},
\end{equation}
where $\phi_i$ is the initial value of the field. The subscript $+$
denotes quantities before the transition and the
 subscript $-$ denotes quantities after the transition.

After the transition, the Klein-Gordon can also be solved despite
the fact that the slow-roll parameter $\eta_-$ becomes temporarily
 large (however $\epsilon_-$ is small), one finds \cite{Martin:2011sn}
\begin{equation}
\phi_-\simeq\phi_0+\frac{\Delta A}{9H_0^2}
\left(1-\left(\frac{\tau}{\tau_0}\right)^3\right)
-\frac{A_-}{3H_0^2}\ln\left(\frac{\tau_0}{\tau}\right),
\end{equation}
where $\Delta A=A_--A_+$ and $\tau_0$ is the transition conformal time.
The slow-roll parameter $\epsilon$ before and after the transition
 can be written respectively as
\begin{equation}
\epsilon_+\simeq\frac{A_+^2}{18M^2_{Pl}H^4_0},
\qquad \epsilon_-\simeq\frac{A_-^2}{18M^2_{Pl}H^4_0}\left(1-\rho^3\tau^3\right)^2,
\end{equation}
where $\rho^3\equiv-({\Delta A}/{A_-})k_0^3$ and $k_0$ is the transition
 scale given by $k_0=a(\tau_0)H_0=-1/\tau_0$. It is worth noting that
before the transition $\eta_+\approx 4\epsilon_+$ and at late times
 (much after the transition) we have again $\eta_-\approx 4\epsilon_-$.
For a transient period after $\tau_0$, $\eta_-$ can become large and
it is well approximated by
\begin{equation}
\eta_-\simeq\frac{6\rho^3\tau^3}{1-\rho^3\tau^3}.
\end{equation}
In Ref.~\cite{Martin:2011sn} it has been shown that the previous
expressions for the slow-roll parameters agree very well with the
exact numerical solutions. For the plots of the next section we
choose the same parameters as the main choice of Ref.~\cite{Martin:2011sn},
 i.e. $V_0=2.37\times10^{-12}M^4_{Pl}$, $A_+=3.35\times10^{-14}M^3_{Pl}$,
 $A_-=7.26\times10^{-15}M^3_{Pl}$, this gives $\Delta A/A_-\approx-3.61$.
 This choice satisfies the constraint on the normalization of the
primordial power spectrum imposed by COBE observations.

We now briefly describe the analytical approximation for the mode
functions, all the details of the derivation can be found
in~\cite{Starobinsky:1992ts,Martin:2011sn}.

We work in the comoving gauge, so that the background value of
$\phi$, i.e. $\phi_0$, determines when the transition happens even
including perturbations. In this gauge the three-dimensional metric
 is perturbed as
\begin{eqnarray}
h_{ij}=a^2e^{2\mathcal{R}}\delta_{ij},
\end{eqnarray}
where $\mathcal{R}$ denotes the comoving curvature perturbation
and tensor perturbations were neglected because they do not contribute
 for the tree-level scalar bispectrum.

The Mukhanov-Sasaki equation in Fourier space is
\begin{equation}
\mathcal{R}_k''+2\frac{z'}{z}\mathcal{R}_k'+k^2\mathcal{R}_k=0,
\label{MSequation}
\end{equation}
where $z=a\sqrt{2\epsilon}$, prime denotes derivative with respect to
the conformal time $\tau$ and the Fourier transform is defined by
\begin{equation}
\mathcal{R}_\mathbf{k}\equiv\mathcal{R}(\tau,\mathbf{k})
=\int d^3x\mathcal{R}(\tau,\mathbf{x})e^{-i\mathbf{k}\cdot\mathbf{x}}.
\end{equation}

The standard quantization procedure in quantum field theory is to
promote $\mathcal{R}$ to an operator that is expanded in terms of
creation and annihilation operators and mode functions as
\begin{eqnarray}
\hat{\mathcal{R}}(\tau,\mathbf{k})
=\mathcal{R}(\tau,\mathbf{k})\hat{a}(\mathbf{k})
+\mathcal{R}^*(\tau,-\mathbf{k})\hat{a}^{\dagger}(-\mathbf{k}),
\end{eqnarray}
where the annihilation operator $\hat{a}$ and the creation operator
$\hat{a}^\dagger$ satisfy the usual commutation relation
$[\hat{a}(\mathbf{k}),\hat{a}^\dagger(\mathbf{k}')]=(2\pi)^3 \delta^{(3)}(\mathbf{k}-\mathbf{k}')$.
The power spectrum of the curvature perturbation is given by
\begin{equation}
2\pi P^{1/2}_\mathcal{R}(k)=\sqrt{2k^3}|\mathcal{R}_k(\tau_f)|,
\label{powerspectrumdef}
\end{equation}
where $\tau_f$ is the end of inflation time.

Before the transition, the slow-roll approximation is valid and the mode
 function solution of Eq.~(\ref{MSequation}) that reduces to
the usual Minkowski result on small scales is the well known leading
order slow-roll expression~\cite{Starobinsky:1992ts,Martin:2011sn},
\begin{equation}
\mathcal{R}_+(\tau,\mathbf{k})
=-\frac{3H_0^3}{A_+}\frac{1}{\sqrt{2k^3}}\left(k\tau-i\right)e^{-ik\tau},
\end{equation}
and one can take the derivative to find
\begin{equation}
\mathcal{R}'_+(\tau,\mathbf{k})
=-\frac{3H_0^3}{A_+}\frac{1}{\sqrt{2k^3}}\left(-ik^2\tau\right)e^{-ik\tau}.
\end{equation}
This last equation is valid at leading order in slow-roll and will be
used in the computation of the dominant contribution to the bispectrum.
After the transition an analytical approximation can also be found,
see~\cite{Starobinsky:1992ts,Martin:2011sn} for all the details.
Here we just present the final result which is given by
\begin{equation}
\mathcal{R}_-(\tau,\mathbf{k})
=\frac{iH_0\alpha_k}{2M_{Pl}\sqrt{k^3\epsilon_-}}\left(1+ik\tau\right)e^{-ik\tau}
-\frac{iH_0\beta_k}{2M_{Pl}\sqrt{k^3\epsilon_-}}\left(1-ik\tau\right)e^{ik\tau},
 \label{modefunctionafter}
\end{equation}
where the Bogoliubov coefficients are
\begin{equation}
\alpha_k
=1+\frac{3i\Delta A}{2A_+}\frac{k_0}{k}\left(1+\frac{k_0^2}{k^2}\right),
\qquad
\beta_k
=-\frac{3i\Delta A}{2A_+}\frac{k_0}{k}\left(1+i\frac{k_0}{k}\right)^2e^{2ik/k_0}.
\end{equation}
The most important feature of the above solution is the presence of the
negative frequency mode which introduces oscillations in the power spectrum.
For the derivative one finds
\begin{eqnarray}
\mathcal{R}'_-(\tau,\mathbf{k})
&=&\frac{iH_0\alpha_k}{2M_{Pl}\sqrt{k^3\epsilon_-}}
\left[-\mathcal{H}\epsilon_-\left(1+ik\tau\right)
-\frac{\mathcal{H}\eta_-}{2}\left(1+ik\tau\right)+k^2\tau\right]e^{-ik\tau}
\nonumber\\
&&-\frac{iH_0\beta_k}{2M_{Pl}\sqrt{k^3\epsilon_-}}
\left[-\mathcal{H}\epsilon_-\left(1-ik\tau\right)
-\frac{\mathcal{H}\eta_-}{2}\left(1-ik\tau\right)+k^2\tau\right]e^{ik\tau},
\end{eqnarray}
where $\mathcal{H}=aH$ is the conformal Hubble parameter.
This expression will also be used in the computation of the dominant
bispectrum contribution in the next section. Because we are only
interested in the dominant contribution we neglect the first term
 inside the square brackets because it is proportional to $\epsilon$
 which is always small in the present model.

\section{The equilateral bispectrum\label{sec:bispectrum}}

In this section we compute the leading-order dominant contribution
 to the bispectrum in the equilateral limit.
After that we compare our result with previous results in the literature.
Finally we discuss the shape and size of the analogue of the non-linearity
parameter in this model.
We leave for Appendix~\ref{FullBispectrum} the lengthy result of
the dominant contribution to the bispectrum that is valid for any
triangular configuration of the momenta.


\subsection{Calculation using the two vertices action\label{subsec:twovertices}}

In order to compute the tree-level scalar bispectrum, we use the in-in
 formalism~\cite{Schwinger:1960qe,Weinberg:2005vy}. For this we need to
find the cubic-order interaction Hamiltonian, see e.g.
Ref.~\cite{Koyama:2010xj} for a review about this procedure.

The third order action after ignoring total derivative terms can be
found for instance
in~\cite{Maldacena:2002vr,Seery:2005wm,Chen:2006nt,Arroja:2011yu}.
It reads (here we follow the notation of \cite{Arroja:2011yu})
\begin{eqnarray}
S_3&=&M_{Pl}^2\int dtd^3x\bigg[
a^3\epsilon^2\mathcal{R}\dot{\mathcal{R}}^2
+a\epsilon^2\mathcal{R}(\partial\mathcal{R})^2-
2a\epsilon\dot{\mathcal{R}}(\partial
\mathcal{R})(\partial \chi) \nonumber \\ &&\qquad\qquad\qquad +
\frac{a^3\epsilon}{2}\frac{d\eta}{dt}\mathcal{R}^2\dot{\mathcal{R}}
+\frac{\epsilon}{2a}(\partial\mathcal{R})(\partial
\chi) \partial^2 \chi +\frac{\epsilon}{4a}(\partial^2\mathcal{R})
(\partial\chi)^2+ 2 \left(\frac{\eta}{4}\mathcal{R}^2
+\tilde{f}(\mathcal{R})\right)\frac{\delta L}{\delta \mathcal{R}}\bigg|_1 \bigg],
\label{cubicaction}
\end{eqnarray}
where we have defined
\begin{eqnarray}
\chi &=& a^2 \epsilon \partial^{-2} \dot{\mathcal{R}}, \qquad
\frac{\delta
L}{\delta\mathcal{R}}\bigg|_1 = a
\left( \frac{d\partial^2\chi}{dt}+H\partial^2\chi
-\epsilon\partial^2\mathcal{R} \right), \\
\tilde{f}(\mathcal{R})&=& \frac{1}{H}
\mathcal{R}\dot{\mathcal{R}}+\frac{1}{4a^2H^2}
\left[-(\partial\mathcal{R})(\partial\mathcal{R})
+\partial^{-2}(\partial_i\partial_j(\partial_i\mathcal{R}\partial_j\mathcal{R}))
\right]
+\frac{1}{2a^2H}\left[(\partial\mathcal{R})(\partial\chi)
-\partial^{-2}(\partial_i\partial_j(\partial_i\mathcal{R}\partial_j\chi))
\right],
\label{redefinition}
\end{eqnarray}
and $\partial^{-2}$ denotes the inverse Laplacian.

In the present model, the slow-roll parameter $\epsilon$ is taken to
be always small after and before the transition, this means inflation
never stops. However both $\eta$ and $\eta'$ can become large temporarily
after the transition. Therefore the dominant contribution to the
bispectrum is expected to come from the following terms in the action,
\begin{eqnarray}
S_3&\supset&M_{Pl}^2\int dtd^3x\bigg[
\frac{a^3\epsilon}{2}\frac{d\eta}{dt}\mathcal{R}^2\dot{\mathcal{R}}
+ 2 \frac{\eta}{4}\mathcal{R}^2\frac{\delta L}{\delta \mathcal{R}}\bigg|_1 \bigg]
+M_{Pl}^2\int dtd^3x \frac{d}{dt}\bigg[
-\frac{\eta a}{2}\mathcal{R}^2\partial^2\chi
\bigg],\label{cubicactiononevertex}
\end{eqnarray}
where the last term is a total time derivative but it is retained
because it is proportional to $\eta$. This is one of the many terms that
appear when simplifying the action to the form~(\ref{cubicaction}) via
integrations by parts~\cite{Collins:2011mz,Arroja:2011yj},
which can be important and cannot a priori be neglected~\cite{Arroja:2011yj}.
After integration by parts the cubic action can be simplified to
\begin{eqnarray}
S_3&\supset&M_{Pl}^2\int dtd^3x\bigg[
-\epsilon\eta a^3\mathcal{R}\dot{\mathcal{R}}^2
-\frac{\epsilon\eta}{2}a\mathcal{R}^2\partial^2\mathcal{R}\bigg],
\label{newvertices}
\end{eqnarray}
from where one easily finds the cubic-order interaction Hamiltonian as
\begin{eqnarray}
H_{int}(\tau) &=&
M_{Pl}^2\int d^3x\bigg[
\epsilon\eta a\mathcal{R}\mathcal{R}'^2
+\frac{\epsilon\eta}{2}a\mathcal{R}^2\partial^2\mathcal{R}\bigg].
\label{Hint3}
\end{eqnarray}
The tree-level three-point correlation function (or bispectrum) at the
time $\tau_e$ after horizon exit is
\begin{equation}
\langle\Omega|\hat{\mathcal{R}}(\tau_e,\mathbf{k}_1)
\hat{\mathcal{R}}(\tau_e,\mathbf{k}_2)
\hat{\mathcal{R}}(\tau_e,\mathbf{k}_3)|\Omega\rangle=
-i\int_{-\infty}^{\tau_e} d\tau a\langle 0|
[
\hat{\mathcal{R}}(\tau_e,\mathbf{k}_1)
\hat{\mathcal{R}}(\tau_e,\mathbf{k}_2)
\hat{\mathcal{R}}(\tau_e,\mathbf{k}_3),{\hat{H}}_{int}(\tau)]
|0\rangle. \label{interaction}
\end{equation}
More explicitly, the bispectrum is
\begin{eqnarray}
\langle\Omega|\hat{\mathcal{R}}(0,\mathbf{k}_1)
\hat{\mathcal{R}}(0,\mathbf{k}_2)
\hat{\mathcal{R}}(0,\mathbf{k}_3)|\Omega\rangle
&\approx&
(2\pi)^3\delta^{(3)}(\mathbf{K})2M_{Pl}^2
\Im\bigg[\mathcal{R}(0,\mathbf{k}_1)\mathcal{R}(0,\mathbf{k}_2)
\mathcal{R}(0,\mathbf{k}_3)
\int_{-\infty}^0d\tau\eta\epsilon a^2\mathcal{R}^*(\tau,\mathbf{k}_1)\times
\nonumber\\
&&\left(2\mathcal{R}'^*(\tau,\mathbf{k}_2)\mathcal{R}'^*(\tau,\mathbf{k}_3)
-k_1^2\mathcal{R}^*(\tau,\mathbf{k}_2)\mathcal{R}^*(\tau,\mathbf{k}_3)\right)\bigg]
+\mathrm{two\,perms.}, \label{bispectrum}
\end{eqnarray}
where $\mathbf{K}\equiv \mathbf{k}_1+\mathbf{k}_2+\mathbf{k}_3$
and ``two perms." denotes two additional permutations of the term displayed,
and in the rest of this work we set $\tau_e\approx0$.

Because the integrand of the previous expression is different before and
 after the transition, here we compute the contribution before the
transition, i.e. the integral from $-\infty$ to $\tau_0=-1/k_0$, separately
from the contribution after the transition, i.e. the integral from
 $\tau_0$ to zero. Obviously the final answer for the bispectrum is the
sum of these two contributions. As we show below, for the parameter
 choice described in the previous section, the contribution after the
 transition is much larger than the contribution before the transition,
so the later one can be ignored.

\emph{The contribution before the transition} (i.e. the integral
 is performed from $-\infty$ to $\tau_0$ only) is
\begin{eqnarray}
&&\langle\Omega|\mathcal{\hat{R}}(\tau_e,\mathbf{k}_1)
\mathcal{\hat{R}}(\tau_e,\mathbf{k}_2)
\mathcal{\hat{R}}(\tau_e,\mathbf{k}_3)|\Omega\rangle_+=
-(2\pi)^3\delta^{(3)}\left(\mathbf{K}\right)
\frac{\eta_+H_0^4}{32M_{Pl}^4\sqrt{\epsilon_+\epsilon_-^3(\tau_e)}}
\frac{1}{(k_1k_2k_3)^3}
\nonumber\\
&&\times\Im\bigg[\left(\alpha_{k_1}-\beta_{k_1}\right)
\left(\alpha_{k_2}-\beta_{k_2}\right)\left(\alpha_{k_3}-\beta_{k_3}\right)
\bigg(\frac{2ik_2^2k_3^2}{k_t}e^{-i\frac{k_t}{k_0}}
\left(1+\frac{k_1}{k_t}+i\frac{k_1}{k_0}\right)
\nonumber\\
&&\qquad\qquad+k_1^2e^{-i\frac{k_t}{k_0}}
\left(k_0-\frac{k_1k_2k_3}{k_tk_0}+i\frac{k_1k_2k_3}{k_t^2}
+i\frac{k_1k_2+k_2k_3+k_1k_3}{k_t}\right)\bigg)
\bigg]+\mathrm{two \,perms.},\label{generalbeforetransition}
\end{eqnarray}
where $k_t\equiv k_1+k_2+k_3$.
In the equilateral limit and after using the explicit expressions
for the slow-roll parameters it simplifies to
\begin{eqnarray}
\langle\Omega|\mathcal{\hat{R}}(0,\mathbf{k})
\mathcal{\hat{R}}(0,\mathbf{k})\mathcal{\hat{R}}(0,\mathbf{k})|\Omega\rangle_+=
-(2\pi)^3\delta^{(3)}
\left(\mathbf{K}\right)\frac{27}{4}\frac{H_0^8A_+}{M_{Pl}^2A_-^3}
\frac{1}{k^6}
\Im\bigg[\left(\alpha_{k}-\beta_{k}\right)^3
e^{-3i\frac{k}{k_0}}\left(2i-\frac{k}{k_0}+\frac{k_0}{k}\right)\bigg].
\end{eqnarray}
In the large scales limit, i.e. $k\ll k_0$, it further simplifies to
\begin{eqnarray}
\langle\Omega|\mathcal{\hat{R}}(0,\mathbf{k})
\mathcal{\hat{R}}(0,\mathbf{k})\mathcal{\hat{R}}(0,\mathbf{k})|\Omega\rangle_+=
(2\pi)^3\delta^{(3)}\left(\mathbf{K}\right)\frac{27}{4}
\frac{H_0^8}{M_{Pl}^2A_+^2}\frac{1}{k^6}
+\mathcal{O}\left(\frac{1}{k_0^2k^4}\right)
.\label{BispectrumTwoVerticesLargeScalesBefore}
\end{eqnarray}
The previous bispectrum is at leading order proportional to $1/k^6$.
This is the scaling expected for a scale invariant model.

On small scales, i.e. when $k_0\ll k$, we get
\begin{eqnarray}
\langle\Omega|\mathcal{\hat{R}}(0,\mathbf{k})\mathcal{\hat{R}}(0,\mathbf{k})
\mathcal{\hat{R}}(0,\mathbf{k})|\Omega\rangle_+
&=&
-(2\pi)^3\delta^{(3)}\left(\mathbf{K}\right)\frac{27}{4}
\frac{H_0^8A_+}{M_{Pl}^2A_-^3}\frac{1}{k^6}
\nonumber\\
&&\times\left[\frac{k}{k_0}\sin\left(\frac{3k}{k_0}\right)
+\left(2-\frac{9\Delta A}{2A_+}\right)\cos\left(\frac{3k}{k_0}\right)
-\frac{9\Delta A}{2A_+}\cos\left(\frac{k}{k_0}\right)
+\mathcal{O}\left(\frac{k_0}{k}\right)
\right],\label{BispectrumTwoVerticesSmallScalesBefore}
\end{eqnarray}
where the first term in square brackets is the leading one and the
others are the sub-leading corrections, one of which oscillates with a
 different ``angular frequency". As one can see, in this limit the
bispectrum scales as $1/k^5$. This represents a linear growth envelope
with respect to the bispectrum of a scale-invariant model. The ``angular
 frequency" of the leading-order term is set by the transition scale
and is $3/k_0$.

The calculation of the \emph{contribution after the transition}
(i.e. the integral from $\tau_0$ to $\tau_e$) is more involved but can
be done analytically. The full result can be found in
Appendix~\ref{FullBispectrum}. In the equilateral limit it simplifies to
\begin{eqnarray}
\langle\Omega|\mathcal{\hat{R}}(0,\mathbf{k})
\mathcal{\hat{R}}(0,\mathbf{k})\mathcal{\hat{R}}(0,\mathbf{k})|\Omega\rangle_-
&=&
-(2\pi)^3\delta^{(3)}\left(\mathbf{K}\right)
\frac{3^6}{2}\frac{H_0^{12}}{A_-^4}\frac{1}{k^9}
\nonumber\\
&&\qquad\times\Re\bigg[\left(\alpha_{k}-\beta_{k}\right)^3\bigg((\alpha_k^*)^3F_1
+(\beta_k^*)^3F_1^*+(\alpha_k^*)^2\beta_k^*F_2
+\alpha_k^*(\beta_k^*)^2F_2^*\bigg)\bigg],
\nonumber\\
\label{BispectrumTwoVertices}
\end{eqnarray}
where the functions $F_1$ and $F_2$ are defined as (for $\tau_e=0$)
\begin{equation}
F_1=\frac{\rho^3i}{2}\Bigg[-1-e^{-3i\frac{k}{k_0}}
\frac{\left(i-\frac{k}{k_0}\right)^2}{\left(1+\frac{\rho^3}{k_0^3}\right)^3}
\left(1+i\frac{k}{k_0}-\left(\frac{k}{k_0}\right)^2
+3\left(\frac{\rho}{k_0}\right)^3+i\frac{3k\rho^3}{k_0^4}
-\frac{\rho^3k^2}{k_0^5}\right)
\Bigg],
\end{equation}
\begin{equation}
F_2=\frac{\rho^3i}{2}\Bigg[3-e^{-i\frac{k}{k_0}}
\frac{\left(i-\frac{k}{k_0}\right)}{\left(1+\frac{\rho^3}{k_0^3}\right)^3}
\left(-3i+\left(\frac{k}{k_0}\right)^3-i\frac{9\rho^3}{k_0^3}
-i\frac{6k^2\rho^3}{k_0^5}+\frac{k^3\rho^3}{k_0^6}\right)
\Bigg],
\end{equation}
where $\rho^3=-(\Delta A/A_)- k_0^3$.
Equation~(\ref{BispectrumTwoVertices}) is one of the main results of this paper.

On large scales, i.e. $k\ll k_0$, it simplifies to
\begin{eqnarray}
\langle\Omega|\mathcal{\hat{R}}(0,\mathbf{k})\mathcal{\hat{R}}(0,\mathbf{k})
\mathcal{\hat{R}}(0,\mathbf{k})|\Omega\rangle_-
&=&
-(2\pi)^3\delta^{(3)}\left(\mathbf{K}\right)
\frac{3^7}{20}\frac{H_0^{12}\Delta A}{A_-A_+^4}\frac{1}{k^6}
\nonumber\\
&&\times\bigg[\left(\frac{k}{k_0}\right)^2
+\left(\frac{k}{k_0}\right)^4\left(\frac{41}{70}-\frac{4A_+}{5A_-}\right)
+\left(\frac{k}{k_0}\right)^6\left(\frac{-38851}{113400}
-\frac{4246A_+}{14175A_-}+\frac{4A_+^2}{25A_-^2}\right)
\nonumber\\
&&\qquad
+\left(\frac{k}{k_0}\right)^8\left(\frac{225409}{3492720}
+\frac{5701A_+}{39690A_-}-\frac{31A_+^2}{2835A_-^2}\right)
+\mathcal{O}\left(\frac{k}{k_0}\right)^{10}\bigg],
\label{BispectrumTwoVerticesLargeScalesAfter}
\end{eqnarray}
where the first term in square brackets is the leading one and we
have included several other sub-leading terms because they will
be used to compare our result with previously known results available
in the literature. On large scales the quantity
$k^6\times\mathrm{bispectrum}$ is proportional to $\frac{k^2}{k_0^2}$
so its amplitude decreases for larger and larger scales until this
 contribution becomes smaller than the contributions coming from the
slow-roll neglected terms in the previous calculation. These contributions
are expected to be small and perhaps unobservable. Some (but not all)
of these contributions were computed in~\cite{Martin:2011sn} and they
turn out to be of the scale-invariant kind (that is the quantity
 $k^6\times\mathrm{ bispectrum}$ is independent of $k$).

On small scales, i.e. when $k_0\ll k$, we get
\begin{eqnarray}
\langle\Omega|\mathcal{\hat{R}}(0,\mathbf{k})
\mathcal{\hat{R}}(0,\mathbf{k})\mathcal{\hat{R}}(0,\mathbf{k})|\Omega\rangle_-
&=&
(2\pi)^3\delta^{(3)}\left(\mathbf{K}\right)
\frac{3^6}{4}\frac{H_0^{12}\Delta A}{A_+^2A_-^3}\frac{1}{k^6}
\nonumber\\
&&\times\left[\frac{k}{k_0}\sin\left(\frac{3k}{k_0}\right)
+\left(\frac{A_-}{A_+}+2
-\frac{9\Delta A}{2A_+}\right)\cos\left(\frac{3k}{k_0}\right)
-\frac{9\Delta A}{2A_+}\cos\left(\frac{k}{k_0}\right)
+\mathcal{O}\left(\frac{k_0}{k}\right)
\right].\nonumber\\
\label{BispectrumTwoVerticesSmallScalesAfter}
\end{eqnarray}
In this limit, the leading order term (first term inside the square brackets)
is a linear growth envelope with oscillations of an ``angular
frequency" $3/k_0$. This linear growth may potentially have important
implications regarding the detectability of this signal. This result is
similar to the one found for instance in Ref.~\cite{Arroja:2011yu}.
Also for the parameters values choice of the previous section and
used in~\cite{Martin:2011sn} one can show that the
contribution~(\ref{BispectrumTwoVerticesSmallScalesAfter}) to
the integral dominates over Eq.~(\ref{BispectrumTwoVerticesSmallScalesBefore}).
In fact the ratio of these amplitudes is proportional to
$\Delta A/(A_+ \epsilon_+)$ which is always large for the case of interest.
Similar result holds on large scales, where
Eq.~(\ref{BispectrumTwoVerticesLargeScalesAfter}) dominates over
Eq.~(\ref{BispectrumTwoVerticesLargeScalesBefore}) (this is true for
scales larger than the transition but not much larger, because
Eq.~(\ref{BispectrumTwoVerticesLargeScalesAfter}) decays quickly as the length
scale increases).

Finally, it is instructive to estimate the range of wavenumbers $\Delta k$
over which large deviations from Gaussianity maybe be expected in this model.
If the sharp transition of the Starobinsky model~(\ref{potential}) is replaced
with a smooth transition with a width in field space $\Delta\phi$ then the
number of $e$-foldings that takes to cross the transition is
$\Delta N\sim H_0/\dot\phi\Delta\phi\sim -V_0/(M^2_{Pl}A_-)\Delta \phi$.
 Because $\Delta k/k_0\sim1/|\Delta N|$ one finds
$\Delta k/k_0\sim (M^2_{Pl}A_-)/(\Delta \phi V_0)$. In the present model
 $\Delta\phi$ is zero which implies that the range of scales affected
is $\Delta k\rightarrow\infty$. In a more realistic model, we expect that
any transition is eventually smooth in a sufficient small scale, this
transition width will introduce a cut-off for the growth of the bispectrum
as described by the previous equation. For scales smaller than this cut-off
scale the previous equation is not valid, instead we expect the amplitude
of the deviations from Gaussianity to go quickly to zero. This expectation
was recently confirmed in a different but somehow related model
(the potential has a step-like feature)~\cite{Adshead:2011jq}.


\subsection{Comparison with previous works\label{subsec:comparison}}

As mentioned in the introduction the bispectrum in this model was
first computed in~\cite{Takamizu:2010xy}. There, the spatial gradient
expansion approach was used at next-to-leading order. Because of this
their result is expected to be correct only on large scales and also
in the limit $A_+\gg A_-$.
In the equilateral limit, the first equation in Eq.~(6.20) of
\cite{Takamizu:2010xy} is
\begin{eqnarray}
&&\langle\Omega|\mathcal{\hat{R}}(0,\mathbf{k})
\mathcal{\hat{R}}(0,\mathbf{k})\mathcal{\hat{R}}(0,\mathbf{k})|\Omega\rangle
=
-(2\pi)^3\delta^{(3)}\left(\mathbf{K}\right)
\frac{3^7}{10}\frac{H_0^{12}\Delta A}{A_-A_+^4}\frac{1}{k^6}
\nonumber\\
&&\times\left[\left(\frac{k}{k_0}\right)^2
+\frac{4}{5}\left(\frac{k}{k_0}\right)^4\left(1-\frac{A_+}{A_-}\right)
+\frac{4}{25}\left(\frac{k}{k_0}\right)^6\left(1-2\frac{A_+}{A_-}
+\frac{A_+^2}{A_-^2}\right)
+\frac{1}{9}\left(\frac{k}{k_0}\right)^8\left(1-2\frac{A_+}{A_-}
+\frac{A_+^2}{A_-^2}\right)\right].
\label{BispectrumMisao}
\end{eqnarray}
Using the fact that $A_+\gg A_-$ the above equation is, up to a
factor of 2, equal to Eq.~(\ref{BispectrumTwoVerticesLargeScalesAfter})
at leading order but it differs in the term proportional to $(k/k_0)^8$.
This extra factor of 2 in Eq.~(\ref{BispectrumMisao}) is due to a double
counting mistake of the permutations
in~\cite{Takamizu:2010xy}\footnote{We thank Yuichi Takamizu for
discussions on this point.}.
We have not found a reason for the discrepancy in the term
proportional to $(k/k_0)^8$.
Finally one can compare the result of \cite{Takamizu:2010xy} for a general
 triangular configuration with Eq.~(\ref{generalaftertransition}) of
Appendix~\ref{FullBispectrum}. Correcting the factor of 2 difference in the
 overall amplitude due to the double counting mistake, we again find
that their result agrees with ours up to a discrepancy in the sub-leading
term of the order of $k^2/k_0^8$.

The second attempt \cite{Martin:2011sn} to compute the bispectrum (equilateral configuration only) for
this model used the in-in formalism just like in the present paper and
even computed analytically some of the sub-leading order corrections
(although not all of them). They found that the dominant bispectrum
contribution on both large and small scales goes as $1/k^6$ and for small
scales the amplitude oscillates with $\cos\left(\frac{3k}{k_0}\right)$.
 As discussed in the previous subsection we found a different result.
On large scale the bispectrum scales as $1/(k_0^2k^4)$ while on small
scales it possesses a linear growth envelope as $1/(k_0k^5)\sin(3k/k_0)$.
We discuss the origin of this difference below, but in short it appears
that Ref.~\cite{Martin:2011sn} missed a term in the calculation that is
proportional to the Dirac delta function.

In Ref.~\cite{Martin:2011sn} an alternative form of the
action\footnote{One is free to use this form of the action or the form
used in the previous subsection or possibly others, the end result is
the same as we shall show below and the two procedures are equivalent.}
given in Eq.~(\ref{cubicactiononevertex}) is used.
Keeping the term proportional to $\epsilon\eta'$ only one finds
\begin{eqnarray}
\langle\Omega|\mathcal{\hat{R}}(\tau_e,\mathbf{k}_1)
\mathcal{\hat{R}}(\tau_e,\mathbf{k}_2)
\mathcal{\hat{R}}(\tau_e,\mathbf{k}_3)|\Omega\rangle
&=&
(2\pi)^3\delta^{(3)}\left(\mathbf{K}\right)iM_{Pl}^2
\mathcal{R}(\tau_e,\mathbf{k}_1)
\mathcal{R}(\tau_e,\mathbf{k}_2)\mathcal{R}(\tau_e,\mathbf{k}_3)
\nonumber\\&&
\int_{-\infty}^{\tau_e}d\tau a^2\epsilon\eta'
\left(\mathcal{R}^*(\tau,\mathbf{k}_1)
\mathcal{R}^*(\tau,\mathbf{k}_2){\mathcal{R}^*}'(\tau,\mathbf{k}_3)
+\textrm{two perms.}\right)+\textrm{c.c.},\label{epsilonetapterm}
\end{eqnarray}
where ``c.c." denotes the complex conjugate.
The action~(\ref{cubicactiononevertex}) contains two further terms,
one proportional to the first order equations of motion, but this term
gives an exactly zero contribution when evaluated on-shell.
The second term is a boundary term. Which should be evaluated at the past
 infinity boundary, this gives zero once we rotate the contour into the
 imaginary plane to select the free vacuum. And it should also be
evaluated at a later time (a time when all scales of interest are well
outside the horizon).
However at such a late time after the transition, slow-roll will have
been restored and again the contribution from this term is sub-dominant.
Therefore one concludes that Eq.~(\ref{epsilonetapterm}) contains the
dominant contribution.

Using the Klein-Gordon equation, $\dot \eta$ can be written as
\begin{equation}
\dot\eta=-\frac{2}{H}\frac{d^2V}{d\phi^2}+\cdots
=-\frac{2}{H}\frac{A_+-A_-}{|\phi'(\tau_0)|}\delta(\tau-\tau_0)+\cdots,
\label{etadot}
\end{equation}
where $\cdots$ denotes other terms without Dirac delta functions, the last
 equality is valid for the Starobinsky model and we have
used $\delta(\phi_0-\phi)=\delta(\tau-\tau_0)/|\phi'(\tau_0)|$.
If we keep only this term in Eq.~(\ref{epsilonetapterm}) the
 integral simplifies to
\begin{eqnarray}
\langle\Omega|\mathcal{\hat{R}}(\tau_e,\mathbf{k}_1)
\mathcal{\hat{R}}(\tau_e,\mathbf{k}_2)
\mathcal{\hat{R}}(\tau_e,\mathbf{k}_3)|\Omega\rangle
&=&
(2\pi)^3\delta^{(3)}\left(\mathbf{K}\right)
\frac{4M^2_{Pl}\epsilon(\tau_0)\left(A_+-A_-\right)
a^3(\tau_0)}{|\phi'(\tau_0)|H(\tau_0)}
\nonumber\\&&
\times\Im\left[\mathcal{R}(\tau_e,\mathbf{k}_1)
\mathcal{R}(\tau_e,\mathbf{k}_2)\mathcal{R}(\tau_e,\mathbf{k}_3)
\mathcal{R}^*(\tau_0,\mathbf{k}_1)
\mathcal{R}^*(\tau_0,\mathbf{k}_2){\mathcal{R}^*}'(\tau_0,\mathbf{k}_3)\right]
\nonumber\\
&&+\textrm{\,two permutations}.
\label{TPFdelta}
\end{eqnarray}
This contribution dominates over the dominant contribution calculated
in~\cite{Martin:2011sn} on small scales and is of the same order of
magnitude for large scales.

For the equilateral configuration, i.e. $k_1=k_2=k_3=k$, the
bispectrum~(\ref{TPFdelta}) with $\tau_e=0$ simplifies to
\begin{equation}
\langle\Omega|\mathcal{\hat{R}}(0,\mathbf{k})\mathcal{\hat{R}}(0,\mathbf{k})
\mathcal{\hat{R}}(0,\mathbf{k})|\Omega\rangle=
(2\pi)^3\delta^{(3)}\left(\mathbf{K}\right)
\left(-\frac{3^6}{4}\right)\frac{\Delta A}{A_+^2A_-^3}
\frac{H_0^{12}}{k^5k_0}\Im
\bigg[
\left(\alpha_k-\beta_k\right)^3\left(1-i\frac{k_0}{k}\right)^2
e^{-3i\frac{k}{k_0}}
\bigg]
.\label{BispectrumDiracEquilateral}
\end{equation}
In the large scale limit, i.e. $k\ll k_0$, this reads
\begin{equation}
\langle\Omega|\mathcal{\hat{R}}(0,\mathbf{k})\mathcal{\hat{R}}(0,\mathbf{k})
\mathcal{\hat{R}}(0,\mathbf{k})|\Omega\rangle=
(2\pi)^3\delta^{(3)}\left(\mathbf{K}\right)\left(-\frac{3^6}{4}\right)
\frac{\Delta A}{A_+^5}\frac{H_0^{12}}{k^6}
\bigg[
1+\left(\frac{k}{k_0}\right)^2\frac{17A_--2A_+}{10A_-}
+\mathcal{O}\left(\frac{k}{k_0}\right)^{4}
\bigg].
\label{BispectrumDiracLargeScales}
\end{equation}
On small scales, i.e. $k_0\ll k$ one finds
\begin{eqnarray}
\langle\Omega|\mathcal{\hat{R}}(0,\mathbf{k})\mathcal{\hat{R}}(0,\mathbf{k})
\mathcal{\hat{R}}(0,\mathbf{k})|\Omega\rangle
&=&
(2\pi)^3\delta^{(3)}\left(\mathbf{K}\right)\frac{3^6}{4}
\frac{\Delta A}{A_+^2A_-^3}\frac{H_0^{12}}{k^6}
\nonumber\\
&&\times\left[\frac{k}{k_0}\sin\left(\frac{3k}{k_0}\right)
+\left(2-\frac{9\Delta A}{2A_+}\right)\cos\left(\frac{3k}{k_0}\right)
-\frac{9\Delta A}{2A_+}\cos\left(\frac{k}{k_0}\right)
+\mathcal{O}\left(\frac{k_0}{k}\right)
\right].\label{BispectrumDiracSmallScales}
\end{eqnarray}
If one calculates $\mathcal{G}/k^3\sim k^6\times\mathrm{bispectrum}$
(the precise definition of $\mathcal{G}$ is given in the next
subsection) then on small scales one finds that
 $\mathcal{G}/k^3\sim \frac{k}{k_0}\sin\left(\frac{3k}{k_0}\right)$
which is the linear growth found in~\cite{Arroja:2011yu}
and in the previous subsection. On large scales
$\mathcal{G}/k^3\sim\mathrm{constant}$.
In order to obtain the final answer for the bispectrum one needs to
add to the above the contribution from all the other terms that was first
calculated in Ref.~\cite{Martin:2011sn}.

Using their Eqs.~(100)-(104) one has that the contribution is
\begin{eqnarray}
\langle\Omega|\mathcal{\hat{R}}(0,\mathbf{k})
\mathcal{\hat{R}}(0,\mathbf{k})\mathcal{\hat{R}}(0,\mathbf{k})|\Omega\rangle
&=&
(2\pi)^3\delta^{(3)}\left(\mathbf{K}\right)\frac{3^7}{4}
\frac{\Delta A}{A_-^5}\frac{H_0^{12}}{k^6}\left(\frac{k_0}{k}\right)^3
\nonumber\\
&&\!\!\!\!\times\left[
\mathcal{A}_1(k)\sin\left(\frac{k}{k_0}\right)
+\mathcal{A}_2(k)\cos\left(\frac{k}{k_0}\right)
+\mathcal{A}_3(k)\sin\left(\frac{3k}{k_0}\right)
+\mathcal{A}_4(k)\cos\left(\frac{3k}{k_0}\right)
\right],\label{BispectrumLeadingMS}
\end{eqnarray}
where
\begin{eqnarray}
\mathcal{A}_1(k)&=&\frac{3A_-^3\Delta A}{4A_+^5}
\left(1+\frac{k^2}{k_0^2}\right)^2\left[9A_-\left(1+\frac{k^2}{k_0^2}\right)
+A_+\left(-9-9\frac{k^2}{k_0^2}+2\frac{k^4}{k_0^4}\right)\right]
\left(\frac{k_0}{k}\right)^6,
\\
\mathcal{A}_2(k)&=&-\frac{3A_-^3\Delta A}{4A_+^5}
\left(1+\frac{k^2}{k_0^2}\right)^2\left[9A_-\left(1+\frac{k^2}{k_0^2}\right)
-A_+\left(9+11\frac{k^2}{k_0^2}\right)\right]\left(\frac{k_0}{k}\right)^5,
\\
\mathcal{A}_3(k)&=&-\frac{A_-^3}{12A_+^5}\bigg[27(\Delta A)^2
\left(1-\frac{k^2}{k_0^2}\right)-27\Delta A\left(5A_--7A_+\right)
\frac{k^4}{k_0^4}-\left(9A_--11A_+\right)^2\frac{k^6}{k_0^6}
\nonumber\\
&&\qquad\qquad
+6A_+\left(-3A_-+5A_+\right)\frac{k^8}{k_0^8}\bigg]\left(\frac{k_0}{k}\right)^6,
\\
\mathcal{A}_4(k)&=&\frac{A_-^3}{12A_+^5}
\bigg[-27A_-^2\left(-3+\frac{k^2}{k_0^2}\right)\left(1+\frac{k^2}{k_0^2}\right)^2
+18A_-A_+\left(1+\frac{k^2}{k_0^2}\right)\left(-9-7\frac{k^2}{k_0^2}
+6\frac{k^4}{k_0^4}\right)
\nonumber\\
&&\qquad\qquad+A_+^2\left(81+153\frac{k^2}{k_0^2}-9\frac{k^4}{k_0^4}
-93\frac{k^6}{k_0^6}+4\frac{k^8}{k_0^8}\right)\bigg]\left(\frac{k_0}{k}\right)^5.
\end{eqnarray}

\begin{figure}[t]
\centering
 \scalebox{.5}
 {\rotatebox{0}{
     \includegraphics*{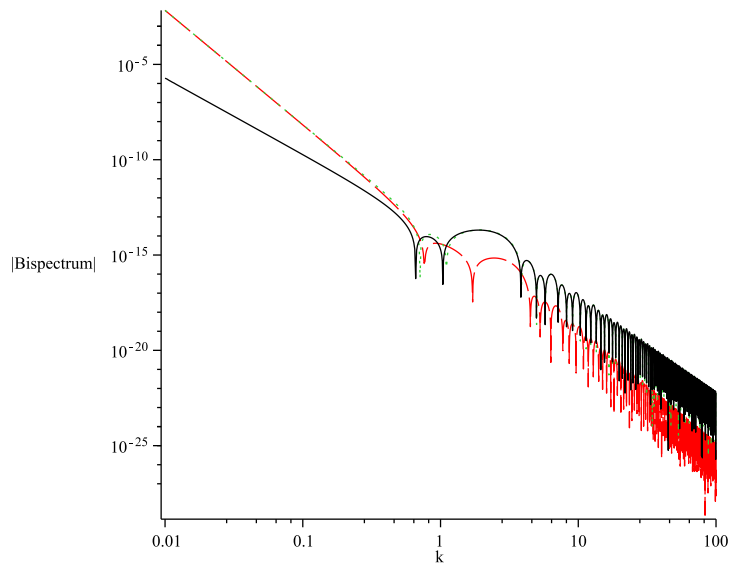}
                 }
 }
\caption{
Plot of the absolute value of the equilateral bispectra for $k_0=1$ as a
function of $k$. The red long-dashed line is the plot of
Eq.~(\ref{BispectrumLeadingMS}), the green dotted line in the plot of the
Dirac delta function contribution to the bispectrum as given by
Eq.~(\ref{BispectrumDiracEquilateral}) and the solid black line is the plot of
Eq.~(\ref{BispectrumTwoVertices}).
One can see that for large scales the contributions from
Eq.~(\ref{BispectrumLeadingMS}) and Eq. (\ref{BispectrumDiracEquilateral})
are comparable in size and because they have opposite signs they partially
 cancel to give the solid black line.
On small scales the amplitude of the long-dashed red line is negligible
compared with the dotted green line. For this scales the dotted green line and
 the solid black line are virtually identical. The model parameters have the
values described in Sec.~\ref{sec:model}.
}
\label{BispectraPlot}
\end{figure}

On large scales, i.e. $k\ll k_0$, one finds that the
bispectrum (\ref{BispectrumLeadingMS}) reads
\begin{equation}
\langle\Omega|\mathcal{\hat{R}}(0,\mathbf{k})
\mathcal{\hat{R}}(0,\mathbf{k})\mathcal{\hat{R}}(0,\mathbf{k})|\Omega\rangle=
(2\pi)^3\delta^{(3)}\left(\mathbf{K}\right)
\frac{3^6}{4}\frac{\Delta A}{A_+^5}\frac{H_0^{12}}{k^6}
\bigg[
1+\left(\frac{k}{k_0}\right)^2\frac{17A_--8A_+}{10A_-}
+\mathcal{O}\left(\frac{k}{k_0}\right)^{4}
\bigg]
.\label{BispectrumLeadingMSLargeScales}
\end{equation}
On small scales, i.e. $k_0\ll k$ one finds
\begin{eqnarray}
\langle\Omega|\mathcal{\hat{R}}(0,\mathbf{k})\mathcal{\hat{R}}(0,\mathbf{k})
\mathcal{\hat{R}}(0,\mathbf{k})|\Omega\rangle
=
(2\pi)^3\delta^{(3)}\left(\mathbf{K}\right)\frac{3^6}{4}
\frac{\Delta A}{A_+^3A_-^2}\frac{H_0^{12}}{k^6}
\left[\cos\left(\frac{3k}{k_0}\right)
+\mathcal{O}\left(\frac{k_0}{k}\right)
\right].\label{BispectrumLeadingMSSmallScales}
\end{eqnarray}
Then it can be seen that the sum of Eq.~(\ref{BispectrumLeadingMS}) with
the Dirac delta function contribution given by
Eq.~(\ref{BispectrumDiracEquilateral}) exactly agrees the
result in Eq.~(\ref{BispectrumTwoVertices}) given in the previous subsection.
This explains the discrepancy between our result and the result
of~\cite{Martin:2011sn} (it also explains the previously existing
discrepancy between the large scales results of \cite{Martin:2011sn}
and \cite{Takamizu:2010xy}).
The point is that the Dirac delta function term in $\dot\eta$ of
Eq.~(\ref{etadot}) cannot be neglected because on large scales it gives
a contribution comparable to Eq.~(\ref{BispectrumLeadingMSLargeScales})
and on small scales it actually dominates over
Eq.~(\ref{BispectrumLeadingMSSmallScales}).
The plots of these different contributions for the final bispectrum
are given in Fig.~\ref{BispectraPlot}, with the choice of model
parameters described in Section~\ref{sec:model}.


\subsection{The non-linearity function
 $\mathcal{G}(k)/k^3$\label{subsec:nonlinearityfunction}}

In this subsection, in order to gain a feeling for the shape and size of
the bispectrum in the equilateral limit for the Starobinsky model,
we plot the previous results for the choice of parameters as in
Section~\ref{sec:model}.

To show the shape of the bispectrum when the three wavenumbers are of
comparable size one commonly plots a quantity defined by
\begin{equation}
\frac{\mathcal{G}(k_1,k_2,k_3)}{k_1k_2k_3}
=\frac{1}{\delta^{(3)}\left(\mathbf{K}\right)}
\frac{(k_1k_2k_3)^2}{(2\pi)^7P_\mathcal{R}^2}
\langle\Omega|\mathcal{\hat{R}}(0,\mathbf{k}_1)
\mathcal{\hat{R}}(0,\mathbf{k}_2)\mathcal{\hat{R}}(0,\mathbf{k}_3)|\Omega
\rangle.
\end{equation}
If this quantity is of order unity there may be some hope that it will
be measured in the near future for example with ESA's Planck satellite.
Because in the present model the amplitudes of the power spectrum before
and after $k_0$ can be quite different it makes more sense to define two
functions $\mathcal{G}_<$ and $\mathcal{G}_>$ to study the bispectrum for
scales larger and smaller than the transition scale $k_0$ respectively, as
\begin{eqnarray}
\frac{\mathcal{G}_<(k_1,k_2,k_3)}{k_1k_2k_3}
&=&\frac{1}{\delta^{(3)}\left(\mathbf{K}\right)}
\frac{(k_1k_2k_3)^2}{(2\pi)^7P_<^2}
\langle\Omega|\mathcal{\hat{R}}(0,\mathbf{k}_1)
\mathcal{\hat{R}}(0,\mathbf{k}_2)\mathcal{\hat{R}}(0,\mathbf{k}_3)|\Omega
\rangle_-,
\\
\frac{\mathcal{G}_>(k_1,k_2,k_3)}{k_1k_2k_3}
&=&\frac{1}{\delta^{(3)}\left(\mathbf{K}\right)}
\frac{(k_1k_2k_3)^2}{(2\pi)^7P_>^2}
\langle\Omega|\mathcal{\hat{R}}(0,\mathbf{k}_1)
\mathcal{\hat{R}}(0,\mathbf{k}_2)\mathcal{\hat{R}}(0,\mathbf{k}_3)|\Omega
\rangle_-,
\end{eqnarray}
where $P_<$ and $P_>$ denote the asymptotic values of the power spectrum
from large and small scales respectively, and are given by
\begin{eqnarray}
P_<&=&\left(\frac{H_0}{2\pi}\right)^2\left(\frac{3H_0^2}{A_+}\right)^2,
\\
P_>&=&\left(\frac{H_0}{2\pi}\right)^2\left(\frac{3H_0^2}{A_-}\right)^2.
\end{eqnarray}
With this last definition we make sure that the oscillations seen
in the quantity $\mathcal{G}_>$ are truly bispectrum oscillations and are
not due to the oscillations already present in the small scales power spectrum.

\begin{figure}[t]
\centering
 \scalebox{.35}
 {\rotatebox{0}{
    \includegraphics*{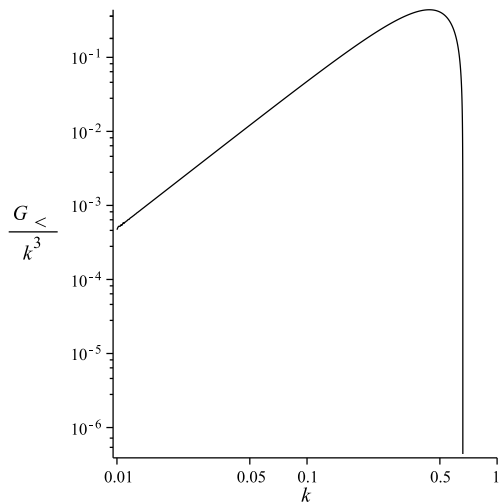}
                 }
 }
\caption{
Plot of $\mathcal{G}_</k^3$ as a function of $k$ for $k_0=1$ and the model
parameters of Sec.~\ref{sec:model}.
On large scales one finds that the amplitude of $\mathcal{G}_</k^3$ is much
smaller than one and is decreasing for larger scales.
This is an expected result as for these scales the transition happened
when they were super-horizon, because $\mathcal{R}$ is non-linearly conserved
the deviation from Gaussianity should be small as these perturbations were
nearly Gaussian at horizon crossing.
For very large scales, eventually the magnitude of this contribution becomes
 smaller than the always present slow-roll corrections that we neglected in
this work. When this happens these sub-leading corrections become the
dominant part. However they are slow-roll suppressed and so less interesting
observationally.
}\label{Gdividedby3ksLSPlot}
\end{figure}

\begin{figure}[t]
\centering
 \scalebox{.7}
 {\rotatebox{0}{
    \includegraphics*{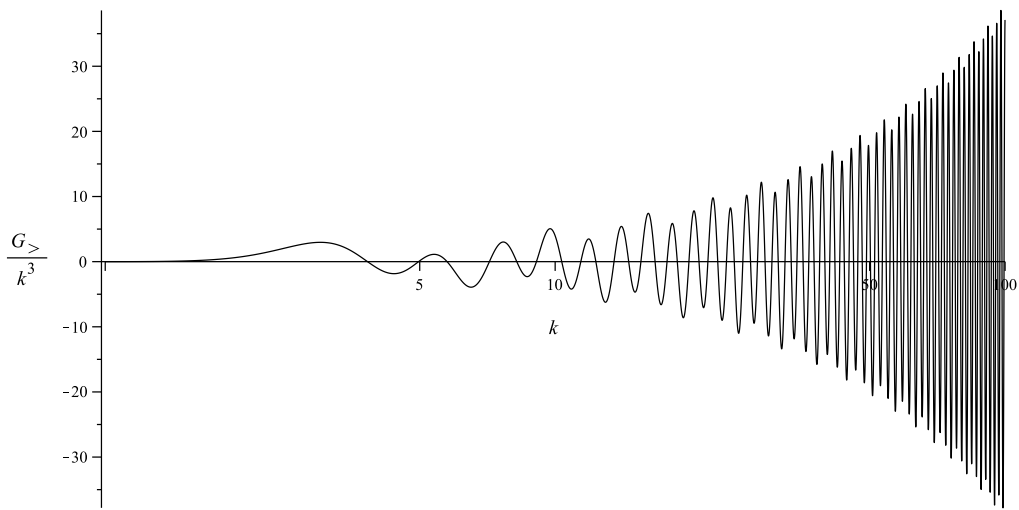}
                 }
 }
\caption{
Plot $\mathcal{G}_>/k^3$ as a function of $k$ for $k_0=1$ and the model
parameters of Sec.~\ref{sec:model}.
On small scales the amplitude of the analogue of the
non-linearity parameter, i.e. $\mathcal{G}_>/k^3$,
oscillates with an ``angular frequency" given by $3/k_0$, where $k_0$
is the transition scale and the envelope of the oscillations grows
 linearly with $k/k_0$.
One easily sees that the amplitude of $\mathcal{G}_>/k^3$ can reach values
much larger than one and be of potential interest observationally.
}\label{Gdividedby3ksSSPlot}
\end{figure}

 As one can see from Fig.~\ref{Gdividedby3ksLSPlot}, on large scales and
with our choice of parameters of Sec. \ref{sec:model} the amplitude of
 $\mathcal{G}_</k^3$ is small and decays quickly towards larger scales.
At some point this amplitude becomes so small that it is subdominant with
respect to the slow-roll correction ignored in this work. Some of these
slow-roll corrections were computed in~\cite{Martin:2011sn}.
On the other hand, Fig.~\ref{Gdividedby3ksSSPlot} shows a very interesting
behaviour for the non-linearity parameter $\mathcal{G}_>/k^3$. One
finds rapid oscillations with a growing amplitude towards smaller scales.
 For the model parameter of Sec.~\ref{sec:model} the amplitude can easily
 become much larger than one.

\section{Conclusion\label{sec:conclusion}}

In this work, we revisited the computation of the tree-level
three-point function (or bispectrum) in the Starobinsky model of inflation.
In this model, a vacuum energy dominated potential is supplemented with a
linear term in which the slope changes abruptly at a point.
It is well known that under the assumption of vacuum energy domination
this model admits an analytical solution not only for the background
equations of motion but also for the mode functions.
Thus despite the fact that the slow-roll approximation is broken for a period
of time just after the transition happens this model still admits an accurate
analytical approximation for the mode functions of linear perturbation theory.

Using this analytical approximation for the mode function we computed
analytically the dominant contribution to the bispectrum.
The main result is Eq.~(\ref{BispectrumTwoVertices}).
In the equilateral limit, we obtained that on scales larger than the scale of the transition
$k_0$ the dominant contribution to the bispectrum scales as $1/(k_0^2k^4)$,
that is, the analogue of
the non-linearity parameter, $\mathcal{G}/k^3$, scales as
 $\mathcal{G}/k^3\sim(k/k_0)^2$. This is in agreement with the leading-order
 result of~\cite{Takamizu:2010xy} (after correcting its factor two error)
 but it differ from the large scale result of~\cite{Martin:2011sn}.
This last result was
 found to scale as $\mathcal{G}/k^3\sim \mathrm{constant}$.
We explained the origin of this difference with the fact that
Ref.~\cite{Martin:2011sn} missed the contribution coming from a term
in the cubic-order action that contains the Dirac delta function
in the time derivative of the coupling constant, $\epsilon\dot\eta$.
This missing contribution has the opposite sign to the dominant term
they calculated and it partially cancels it out and one is left with
the result we found, that is $\mathcal{G}/k^3\sim (k/k_0)^2$.
For small scales, we found some interesting behaviour.
The parameter $\mathcal{G}/k^3$ scales as $\mathcal{G}/k^3\sim k/k_0\sin(3k/k_0)$.
 The ``angular frequency" of the small scale oscillations is set by the
transition scale and is given by $3/k_0$.
The envelope function of these oscillations in a linear growth toward small
scales. A similar result was found in~\cite{Arroja:2011yu} for a somehow
related model.
This linear growth represents a strong scale dependence of the non-linearity
parameter in this model and the $k/k_0$ enhancement factor can have
important consequences regarding observations.
It would be interesting to consider observational constraints on this
kind of oscillating and growing amplitude $\mathcal{G}/k^3$ models,
 however that study is beyond the scope of the present work.
Once again this result differs from the result found in
Ref.~\cite{Martin:2011sn}. There, it was found that $\mathcal{G}/k^3$
scales as $\mathcal{G}/k^3\sim \cos(3k/k_0)$, with no enhancement factor.
The reason for the difference with respect to our result is again the
fact that the contribution from the Dirac delta function was missed.
Finally, we briefly showed that the amplitude of $\mathcal{G}/k^3$ can reach
values much larger than one on small scales. The large enhancement factor
$k/k_0$ is key in reaching that conclusion.

For future work, we leave the computation of the trispectrum or
four-point function.


\begin{acknowledgments}
We would like to thank Yuichi Takamizu for discussions regarding
the comparison of our results with the result of
Takamizu \emph{et al.} \cite{Takamizu:2010xy}.
FA acknowledges the support by the World Class University grant
no. R32-10130 through the National Research Foundation, Ministry of Education,
Science and Technology of Korea.
MS is supported in part by JSPS Grant-in-Aid for Scientific Research (A)
No.~21244033.
This work was also supported in part by MEXT Grant-in-Aid for the global
COE program at Kyoto University, ``The Next Generation of Physics, Spun
from Universality and Emergence".
\end{acknowledgments}

\appendix
\section{\label{FullBispectrum}The bispectrum for any configuration of the triangle}

In this Appendix we will present the result for the bispectrum that is valid
for any allowed configuration of the three momentum vectors.

As discussed in the main text, the integral in Eq.~(\ref{bispectrum}) can be
split in a contribution before the transition and a contribution
after the transition. For a general triangle configuration, the contribution
before the transition is given by Eq.~(\ref{generalbeforetransition}) and it
is sub-leading with respect to the contribution
after the transition. Therefore for the scales of interest the bispectrum is well approximated by the contribution after the transition only. It turns out that the integral after the transition can
also be evaluated analytically. After a lengthy but straightforward calculation,
 the result for the bispectrum is
\begin{eqnarray}
&&\langle\Omega|\mathcal{\hat{R}}(\tau_e,\mathbf{k})
\mathcal{\hat{R}}(\tau_e,\mathbf{k})
\mathcal{\hat{R}}(\tau_e,\mathbf{k})|\Omega\rangle_-=
-(2\pi)^3\delta^{(3)}\left(\mathbf{K}\right)
\frac{9H_0^{6}}{8\sqrt{2}A_-M_{Pl}^3\sqrt{\epsilon_-^3(\tau_e)}}
\frac{1}{(k_1k_2k_3)^3}
\nonumber\\
&&\quad\times\Re\bigg[\left(\alpha_{k_1}-\beta_{k_1}\right)
\left(\alpha_{k_2}-\beta_{k_2}\right)\left(\alpha_{k_3}-\beta_{k_3}\right)
\bigg(
\alpha_{k_1}^*\alpha_{k_2}^*\alpha_{k_3}^*T(k_1,k_2,k_3)
-\beta_{k_1}^*\beta_{k_2}^*\beta_{k_3}^*T(-k_1,-k_2,-k_3)
\nonumber\\
&&\qquad\qquad\qquad
-\alpha_{k_1}^*\alpha_{k_2}^*\beta_{k_3}^*T(k_1,k_2,-k_3)
-\alpha_{k_1}^*\alpha_{k_3}^*\beta_{k_2}^*T(k_1,-k_2,k_3)
-\alpha_{k_2}^*\alpha_{k_3}^*\beta_{k_1}^*T(-k_1,k_2,k_3)
\nonumber\\
&&\qquad\qquad\qquad
+\alpha_{k_1}^*\beta_{k_2}^*\beta_{k_3}^*T(k_1,-k_2,-k_3)
+\alpha_{k_2}^*\beta_{k_3}^*\beta_{k_1}^*T(-k_1,k_2,-k_3)
+\alpha_{k_3}^*\beta_{k_1}^*\beta_{k_2}^*T(-k_1,-k_2,k_3)
\bigg)\bigg],
\label{generalaftertransition}
\end{eqnarray}
where $T(k_1,k_2,k_3)$ is defined as
\begin{eqnarray}
T(k_1,k_2,k_3)&=&\bigg[
\frac{i\rho^3}{(1-\rho^3\tau^3)^3}e^{i\tau k_t}
\bigg(
-3+k_1k_2k_3k_t\rho^3\tau^7
+i\rho^3\tau^6\left(9k_1k_2k_3+k_1(k_2^2+k_3^2)+k_2(k_1^2+k_3^2)
+k_3(k_1^2+k_2^2)\right)
\nonumber\\
&&\quad
-\rho^3\tau^5\left(k_1^2+k_2^2+k_3^2+9(k_1k_2+k_2k_3+k_3k_1)\right)
-\tau^4\left(k_1k_2k_3k_t+9i\rho^3k_t\right)
\nonumber\\
&&\quad
+\tau^3\left(9\rho^3-ik_t(k_1k_2+k_2k_3+k_3k_1)\right)
+\tau^2\left(k_1^2+k_2^2+k_3^2+3(k_1k_2+k_2k_3+k_3k_1)\right)
+3ik_t\tau
\bigg)
\bigg]\bigg|_{-k_0^{-1}}^{\tau_e}.\nonumber\\
\end{eqnarray}
For $\tau_e=0$, the function $T(k_1,k_2,k_3)$ is related to the functions $F_1$ and $F_2$
defined in the main text as $F_1=T(k,k,k)/6$ and $F_2=-T(k,k,-k)/2$ and in the equilateral configuration Eq.~(\ref{generalaftertransition}) reduces to Eq.~(\ref{BispectrumTwoVertices}) .

Finally it is worth noting that in the squeezed limit of the triangle formed
 by the three momenta, i.e. when $k_2\sim k_3\sim k$ and the small momentum
 $k_1$ is $k_1\ll k$, one can find on large scales, i.e. when $k_0\gg k\gg k_1$,
\begin{eqnarray}
\langle\Omega|\mathcal{\hat{R}}(0,\mathbf{k})
\mathcal{\hat{R}}(0,\mathbf{k})\mathcal{\hat{R}}(0,\mathbf{k})|\Omega\rangle_-
\approx
-(2\pi)^3\delta^{(3)}\left(\mathbf{K}\right)\frac{162}{5}
\frac{H_0^{12}\Delta A}{A_-A_+^4}\frac{1}{k_1^3k^3}\left(\frac{k}{k_0}\right)^2,
\end{eqnarray}
and on small scales, i.e. when $k_0\ll k_1\ll k$, one finds
\begin{eqnarray}
\langle\Omega|\mathcal{\hat{R}}(0,\mathbf{k})
\mathcal{\hat{R}}(0,\mathbf{k})\mathcal{\hat{R}}(0,\mathbf{k})|\Omega\rangle_-
\approx
(2\pi)^3\delta^{(3)}\left(\mathbf{K}\right)\frac{3^5}{2}
\frac{H_0^{12}\Delta A}{A_-^3A_+^2}\frac{1}{k_1^3k^3}
\left(\frac{k_1}{k_0}\sin\left(\frac{2k}{k_0}\right)\right),
\end{eqnarray}
where the enhancement factor $k_1/k_0$ in the previous equation is $k_1/k_0\gg 1$.


\end{document}